,

# A Novel Approach For Intranet Mailing For Providing User Authentication

**ASN Chakravarthy**[†]
Associate Professor, Dept. of CSE
Sri Sai Aditya Institute of Science & Technology
Suram Palem, E.G.Dist, Andhra Pradesh, India

**A.S.S.D.Toyaza**[††]
Final Year M.Tech, Dept. of CSE,
Sri Sai Aditya Institute of Science & Technology,
Suram Palem, E.G.Dist, Andhra Pradesh, India

**Summary**
With the explosion of the public Internet and e-commerce, private computers, and computer networks, if not adequately secured, are increasingly vulnerable to damaging attacks. Hackers, viruses, vindictive employees and even human error all represent clear and present dangers to networks. Various antidotes that are in fact inextricable with security issues are – Cryptography, Authentication, Integrity and Non Repudiation, Key Distribution and certification, Access control by implementing Firewalls etc. The main idea of this paper is to overcome the PGP's(Pretty Good Privacy) main limitation of incomplete non-repudiation Service, which increases the degree of security and efficiency of an email message communication through NRR(Non-Repudiation of Receipt) and including PGPs original feature of NRO(Non-Repudiation of Origin), and there it assures new security service of Mutual Non-Repudiation (MNR)
.

## 1. Introduction

Non-repudiation service can be viewed as an extension to the identification and authentication service. In general, non-repudiation applies when data is transmitted electronically; for example, an order to a stock broker to buy or sell stock, or an order to a bank to transfer funds from one account to another. The overall goal is to be able to prove that a particular message is associated with a particular individual. Non-repudiation is the assurance that someone cannot deny something. Typically, non-repudiation refers to the ability to ensure that a party to a contract or a communication cannot deny the authenticity of their signature on a document or the sending of a message that they originated.

Repudiation of deliver occurs when the sender claims to have sent the message, but the recipient denies receiving it; the sender claims to have received; or the sender and the recipient claim different date or time of receiving the message. The repudiation of delivery could be triggered by the same events as the repudiation of origin; misinformation, lying. Communication error , or a third-party intervention.

Authentication is the act of confirming the truth of an attribute of a datum or entity. This might involve confirming the identity of a person, tracing the origins of an artifact, ensuring that a product is what it's packaging and labeling claims to be, or assuring that a computer program is a trusted one. The authentication of information can pose special problems (especially man-in-the-middle attacks), and is often wrapped up with authenticating identity. Literary can involve imitating the style of a famous author. If an original manuscript, typewritten text, or recording is available, then the medium itself (or its packaging - anything from a box to e-mail headers) can help prove or disprove the authenticity of the document.

With the growing use of the Internet as a medium for doing business, purchasing products, and exchanging personal and private information, the need for a secure and verifiable mechanism for information transfer and exchange is becoming critical. One of the biggest difficulties since the inception of the e-mail message communication over an open network is providing security to email communication. Many protocols have been developed to provide security and authentication for the e-mail message. Some of the protocols are Simple Mail Transfer Protocol (SMTP), Multipurpose Internet Mail Extension (MIME), and its enhancement, known as Secure MIME (S/MIME). Other protocols are: Certified Exchange of Electronic Mail (CEEM), Secure E-mail Protocol (SEP), Privacy Enhanced Mail (PEM) and Pretty Good Privacy (PGP). Among these protocols PGP is one of the secured and enhanced protocol.

1.1 Existing System

Pretty Good Privacy (PGP) is a popular program used to encrypt and decrypt email over the Internet. It can also be used to send an encrypted digital signature that lets the receiver verify the sender's identity and know that the message was not changed in the transmission. PGP is available both as freeware and in a low-cost commercial version, is the most widely used privacy-ensuring program by individuals and is also used by many corporations. Developed by Philip R. Zimmermann in 1991, PGP has become a standard for e-mail security. PGP can also be used to encrypt files being stored so that they are unreadable by other users or intruders, PGP can be used basically for 4 things [1]:
• Encrypting a message or file so that only the recipient can decrypt and read it. The sender, by digitally signing with PGP, can also guarantee to the recipient, that the message or file must have come from the sender and not an impostor.
• Clear signing a plain text message guarantees that it can only have come from the sender and not an impostor.
• Encrypting computer files so that they can't be decrypted by anyone other than the person who encrypted them.
• Really deleting files (i.e. overwriting the content so that it can't be recovered and read by anyone else) rather than just removing the file name from a directory/folder.
PGP provides two services: encryption and digital signatures

1.2 Proposed System

Enhanced Pretty Good Privacy (EPGP) is a new cryptosystem based on Pretty Good Privacy (PGP), used for the purpose of secure e-mail message communication over an open network. The idea of EPGP, in this paper is to overcome PGP's main drawback of incomplete non-repudiation service, and therefore, attempts to increase the degree of security and efficiency of e-mail message communication through the concept of NRR, plus PGP's original feature of NRO, and therefore, assuring the new security service of *Mutual* Non-Repudiation (MNR) for an e-mail message communication.

"Non-Repudiation of Receipt"(NRR), is a cryptographic method that makes sure that the sender of information is protected against the denial of the receiver, who may say the sender never sent the information, or that he didn't send it on time. With NRR, the sender saves the digitally signed message he sent and when receiving the message, the receiving party must extract the message, digitally sign it and then send it back to the sender. NRR provides legal evidence that the denying party did receive the information by using digital signatures for proof.

NRR can also be defined as a service that provides proof of the integrity and origin of data, both in an unforgivable relationship, which can be verified by any third party at any time; or, an authentication that with high assurance can be asserted to be genuine, and that cannot subsequently be reputed.

"Non-Repudiation of Origin"(NRO), is a cryptographic method that makes sure that the original sender of information cannot successfully deny that he sent the information because it can be verified that he had sent it. NRO provides legal evidence that the denying party sent the information by using digital signatures for proof. Non-repudiation of origin defines requirements to provide evidence to users/subjects about the identity of the originator of some information. The originator cannot successfully deny having sent the information because evidence of origin (e.g. digital signature) provides evidence of the binding between the originator and the information sent. The recipient or a third party can verify the evidence of origin. This evidence should not be forgeable.

We have replaced the **LZ77** algorithm with deflator and enflaltor algorithm for compressing and decompressing these algorithms are combination of **LZW + Huffman coding**. Then in 4th phase we have replaced the **DES_CBC** symmetric encryption algorithm with the **tripleDES** algorithm

## 2. EPGP Algorithm

EPGP has solved the problem of providing the security and authentication and provides complete fair and non-repudiation service for the email message.
E-mail communication process is a connectionless-oriented type of communication in which it is necessary for both sides of communication to be in direct contact with each other simultaneously during the transmission and reception phases.

Instead, an e-mail message $M_5$ that sender A sends is uploaded to a 24-hour-available trusted e-mail software server D. Then whenever receiver B opens its e-mail inbox, message $M_5$ is downloaded from e-mail server D to B's machine, where B's email software performs the reverse PGP process to retain back the original text of e-mail message M.
It is not necessary for B to be online when A sends the message, neither is it necessary for A to be online when B receives the message, since the email server D is online all the time. Server D is *not* a Trusted Third Party (TTP) from outside the communication link, but it is an embedded part in the whole process that takes on the role of message delivery.

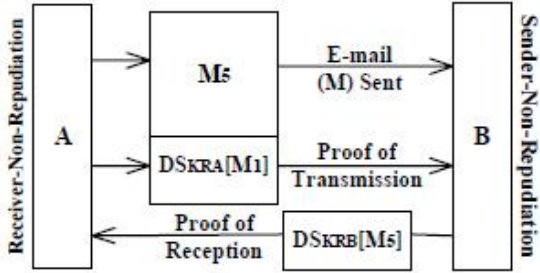

Figure (1): Fair MNR of EPGP

The entire EPGP process consists of three main phases, described as follows:
1. Transmission Phase
2. NNR Phase
3. Reception Phase

## 3. Implementation

Phase I: This is "transmission phase", where similar steps like PGP will be taken place. User A's e-mail software computes a message $M_1$, by hashing message M, using the SHA-1 hashing algorithm as follows:

A: $M_1 = H(M)$

Then, user A's e-mail software computes $M_2$ as a digital signature of message $M_1$, using the DSS digital signature scheme. The attached digital signature of sender A, $DS_{KRA}[M_1]$, on to the message will assure the feature of NRO, which is already achieved by PGP as well, as follows:

A: $M_2 = DS_{KRA}[M_1] \parallel M$

Then user A's email software compresses message $M_2$ as message $M_3$, using the LZW + Huffman coding algorithm of deflator zipping as follows,

A: $M_3 = Z(M_2)$

Then, user A's e-mail software computes $M_4$, by encrypting message $M_3$, by the secret key $K_S$, using a DES-CBC symmetric encryption algorithm.

A: $M_4 = E_{Ks}[M_3]$

Finally, user A's e-mail software computes $M_5$ by applying Radix-64 conversion to ASCII on message $M_4$, and sends the final message to e-mail server D, as follows:

A→D: $M_5 = R64(M_4)$

Now, the message has been sent to receiver B via server D over the open network. It is clear now that receiver B till now is still not able to decrypt the message since it has not gotten yet the secret key $K_S$, nor server D's private key, $K_{RD}$. The enhancement of NNR is applied here as shown in the next phase of the EPGP process. The entire "transmission phase" of EPGP is illustrated in figure (2).

Phase II: This is called "NNR phase", which is the major enhancement of EPGP. Once receiver B, opens its e-mail inbox, downloads message $M_5$ from server D, and attempts to open message M, user B's e-mail software will establish a communication session with server D to get the secret key, $K_S$, to decrypt the message. First of all, server D forwards message $M_5$ to B, as follows:

D→B: $M_5$

Server D will not grant receiver B the secret key, $K_S$, unless receiver B handles its digital signature on the unopened message, M5, to server D first. This will serve as evidence of message reception, and therefore, the MNR of the whole process. Receiver B submits server D its digital signature on the received message, M5, encrypted by RSA, using user A's public key, KUA, as follows:

B→D: $M_6 = E_{KUA}[DS_{KRB}[M_5]]$

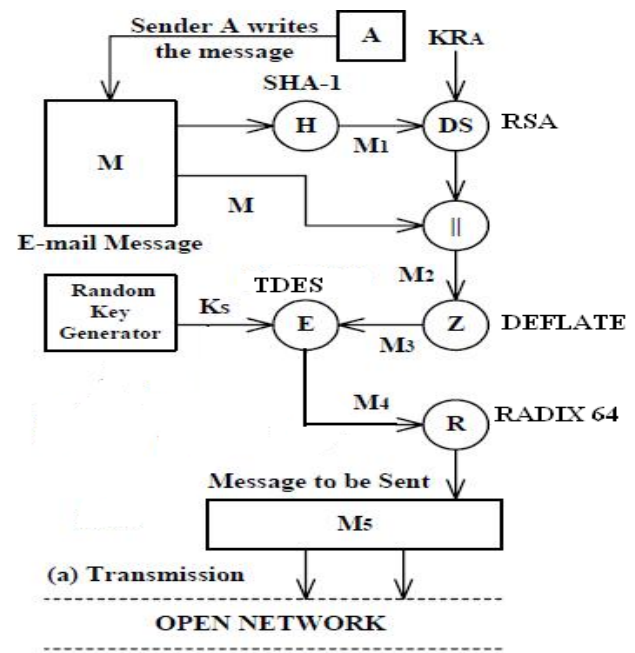

Figure (2): The EPGP Transmission Phase

Then, server D may send the secret key, KS, to receiver B. Now, server D performs its last task by simply forwarding user B's digital signature on message $M_5$ to user A, as follows:

D→A: $M_6 = E_{KUA}[DS_{KRB}[M_5]]$

Now, the main objective of Mutual Non-Repudiation (MNR) of the whole e-mail communication service is finally achieved, and receiver B can no more deny receiving M, since A can prove such reception, as follows:

A: $DS_{KRB}[M_5] = D_{KRA}[M_6]$
$= D_{KRA}[E_{KUA}[DS_{KRB}[M5]]]$

Now, receiver B can finally get the needed secret key, KS, to decrypt the e-mail message and obtain the original text of the transmitted e-mail message M, sent by sender A, as follows:

**B:** $K_S = D_{KRB}[E_{KUB}[K_S]]$

Receiver B is now ready to decrypt the whole email message, as shown in the next phase of the EPGP process. The entire "NNR phase" of EPGP, which assures the MNR of the system, is illustrated in figure (3).

Phase III: This is the "reception phase". It is totally similar to PGP. Receiver B, who got the secret key KS, retains back the original email message M from the received message $M_5$, in the same procedure of the PGP process, as follows:

**B:** $M_4 = R_{64}^{-1}(M_4)$
$= E_{KS}[M_3] \parallel E_{KUB}[K_S]$

**B:** $K_S = D_{KRB}[E_{KUB}[K_S]]$ (using triple DES)

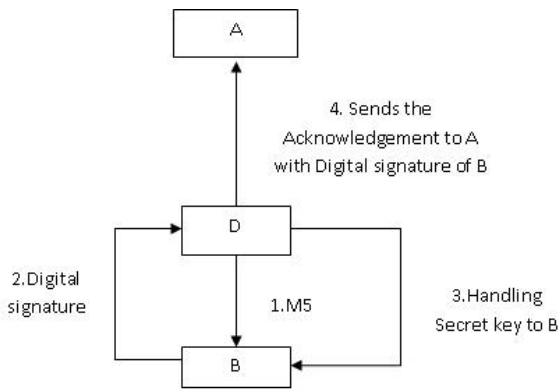

Figure (3): The EPGP NNR Phase

**B:** $M_3 = D_{KS}[E_{KS}[M_4]]$

**B:** $M_2 = Z_{-1}(M_3)$
$= DS_{KRA}[M_1] \parallel M^*$

**B:** $M_1 = DS^{-1}_{KUA}[DS_{KRA}[M_1]]$

**B:** IF $H(M^*) = H(M)$

**THEN $M^* = M$**

Finally receiver B has retained back the original email message, M, which was sent by sender A, and the whole EPGP process is now complete.

## 4. Experiment & Results

In order to have an intranet mailing each and every user must have a login id and password which can be created if they enter their authenticated details. So, when the user enters the details his email id becomes his login id and on his mail id a separate key will be created in the database which will be useful for the future purpose figure(4). On Clicking the submit button the page will be forwarded to the login screen.

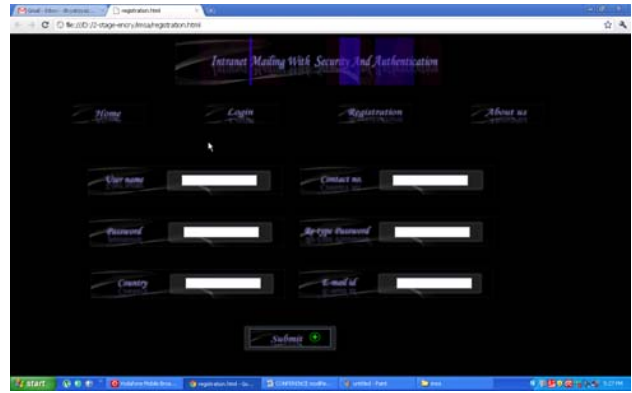

Figure (4): Registration For Users

The login id and password are enough to get into this. If the user forgets his password or id he can get back the details through a mail. After logging in, the user can get a preview of his inbox where he can perform different actions like delete, reading of the mail, forwarding, replying etc., the Figure(5) will illustrate us the login for users.

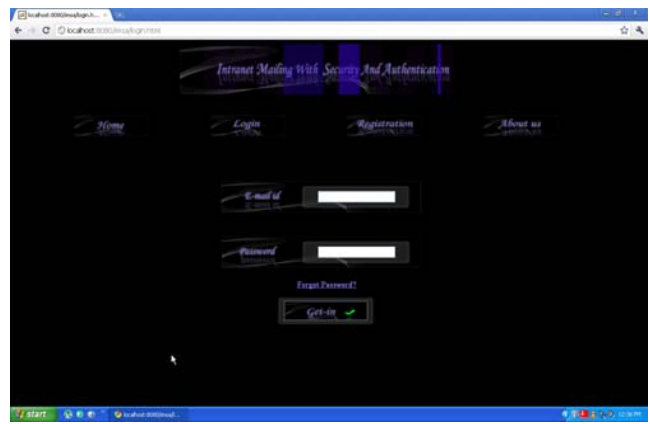

Figure (5): Login For User

The user after logging in have to compose a mail as in figure (6) which will be transferred to the specified email id in the column of To, in the encrypted format. And the user will click on the send button. Now the background process of the EPGP algorithm as described above will be taken place.

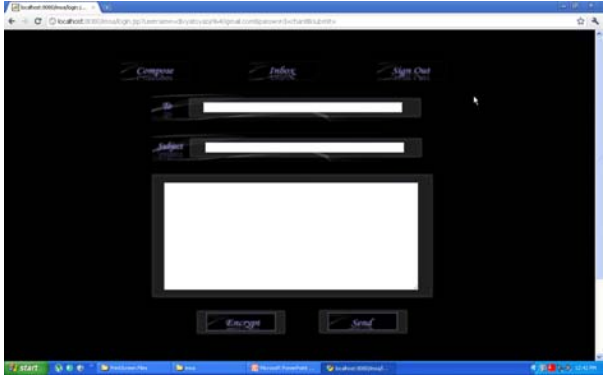

Figure (6): Composing The Mail

The message after composing will be sent to the server in the encrypted format which is in figure (7). The encrypted form if the message will be sent to the server D which is displayed as message sent. Along with the encrypted message secret key and the private key must be sent to the receiver in order to decrypt the message.

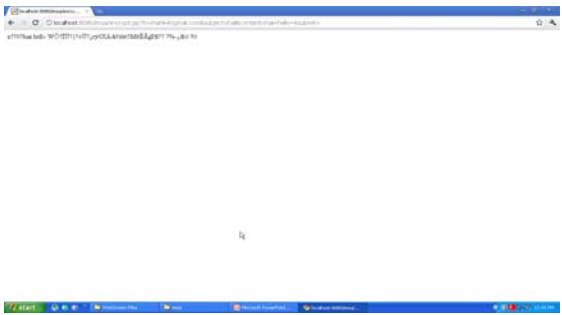

Figure (7): Encrypted Form of Message

For, the receiver B email will be displayed in the inbox but he cannot view the content of the mail. He can do so if he downloads the secret key and the private key form the server D. the same process of logging in and viewing must be followed as sender A.

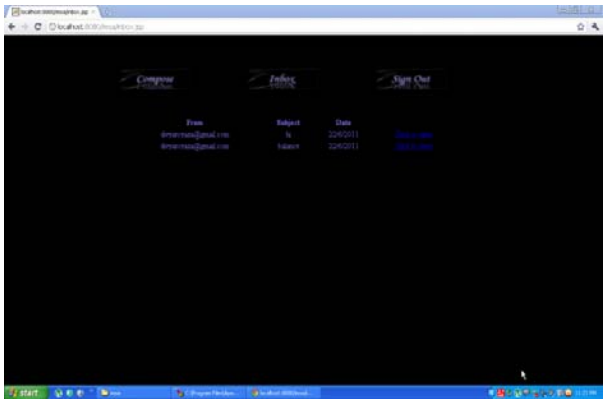

Figure (8): Inbox preview of Mail

For the receiver B inbox preview will be as shown in figure (8). From, subject and date fields will be displayed and link to preview that message is displayed in the page.

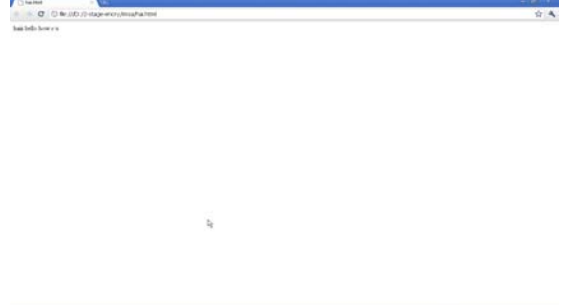

Figure (9): Decrypted Form Of Mail

The comparison graph between the PGP and the EPGP is as followed in the figure (10), which illustrates that the EPGP works efficient and secured and authenticated than PGP.

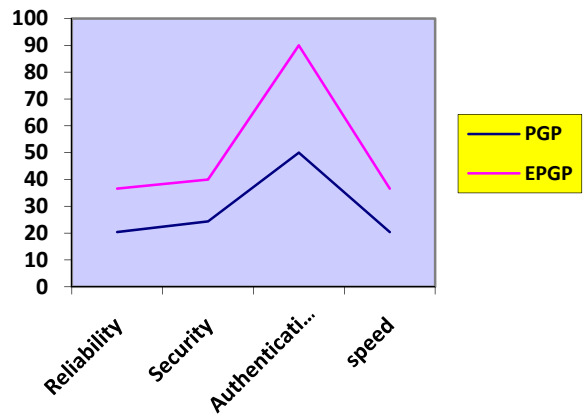

Figure (10): Comparison of PGP and EPGP.

## 5. Conclusion & Future Scope

The current state of password storage is enough for now, although may not be enough in the future. The Blowfish algorithm, having been proven untraceable for the 10 years of its lifetime, seems to be the most secure hash algorithm for passwords, and will remain so for the foreseeable future. MD4 is a provably weak algorithm, susceptible to a two-round attack, and MD5 suffers from this weakness as well. Authentication methods are currently very good, but may not be enough for too much longer, as computers get more powerful. A possible replacement to the MD5 hash is the SHA-1 algorithm, which, as previously stated, offers 160 bits of encryption. However, Blowfish is a currently used algorithm on FreeBSD and OpenBSD, very secure, and can very easily be ported to other UNIX-like operating systems. This makes it a very powerful choice for future directions of password storage. In this paper we introduced a new protocol with highly secured architectural design by adding non-repudiation feature to PGP. We achieved the mutual non repudiation with implementation of EPGP; with this the security of email message communication is improved. We considered simulation of EPGP for the improvement of EPGP model and to solve drawbacks of PGP by adding or replacing further more complex and secure algorithms to make EPGP secure and efficient for the purpose of secure email message communication. In future we can provide enhanced authentication scheme by using neural network associative memories to replace traditional authentication schemes.

## 6. References

[1] European Intensive Programme on Information and Communication Technologies Security IPICS'99, E-Mail Security: PGP (Pretty Good Privacy) & PEM (Privacy-Enhanced Mail) by Konstantinos Raptis from university of TIE AEGEAN.

[2] Tanenbaum, A , "Computer Networks", Prentice Hall Upper Saddle River, NJ, c1996

[3] Stallings W, Network Security Essentials, Prentice Hall, Upper Saddle River, NJ, c2003.

[4] Michael Roe, "A Technical report on Cryptography and evidence ", UNIVERSITY OF CAMBRIDGE.

[5] Drummond R, Cox N, "Lan Times E-mailResource Guide", Osbone McGraw-Hill, 1994.

[6] Al-Hammadi B and Shahsavari M, "Certified Exchange of Electronic Mail (CEEM)", Southeastcon '99, IEEE Proceedings, 1990, pp. 40-43.

[7] http://en.wikipedia.org/wiki/SHA-1

[8] http://en.wikipedia.org/wiki/DEFLATE

[9] http://en.wikipedia.org/wiki/Base64

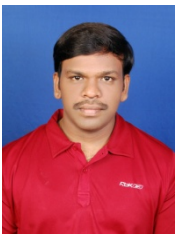
[†]**A.S.N Chakravarthy** received his M.Tech (CSE) from JNTU, Anantapur, Andhra Pradesh, India and pusuing PhD in Network Security from Acharya Nagarjuna Univearsity, Guntur, Andhra Pradesh, India. Presently he is working as an Associate Professor in Computer Science and Engineering in Sri Aditya Institute of Science & Technology, SuramPalem, E.G.Dist, AP, India. His research area includes Network Security, Cryptography, Intrusion Detection, Neural networks.

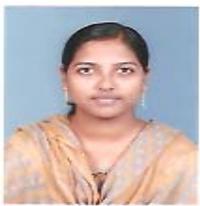
[††]**A.S.S.D.Toyaza r**eceived her B.Tech degree in Computer Science and Engineering from Aditya Engineering College in 2009. Now pursuing M.Tech degree in Computer Science and Engineering from Sri Sai Aditya Institute of Science &Technology, SuramPalem, E.G.Dist, AP, India.